\documentclass[%
 reprint,
 amsmath,amssymb,
 aps,
prl
]{revtex4-1}

\usepackage{graphicx}% Include figure files
\usepackage{dcolumn}% Align table columns on decimal point
\usepackage{bm}% bold math
\def \k {{\bm k}}
\def \el {\varepsilon_{l\sigma}}
\begin{document}

%\preprint{APS/123-QED}

\title{Universal Quantization of Magnetic Susceptibility Jump at Topological Phase Transition}

\author{Soshun Ozaki}
\email{ozaki@hosi.phys.s.u-tokyo.ac.jp}
% \altaffiliation[Also at ]{Physics Department, XYZ University.}%Lines break automatically or can be forced with \\
\author{Masao Ogata}%
% \email{Second.Author@institution.edu}
\affiliation{%
 Department of Physics, University of Tokyo, Bunkyo, Tokyo 113-0033, Japan
}%

\date{\today}% It is always \today, today,
             %  but any date may be explicitly specified

\begin{abstract}
We examine the magnetic susceptibility of topological insulators microscopically
and find that the orbital--Zeeman (OZ) cross term, 
the cross term between the orbital effect and 
the spin Zeeman effect, is directly related to the Berry curvature when 
the $z$-component of spin is conserved.
In particular,
the OZ cross term reflects the spin Chern number,
which results in the quantization of the magnetic susceptibility jump at the topological phase transition.
The magnitude of the jump is in units of the universal value $4|e|\mu_{\rm B}/h$.
% with $\mu_{\rm B}$ being the Bohr magneton.
The physical origin of this quantization is clarified.
We also apply the obtained formula to an explicit model and demonstrate the quantization.
\end{abstract}

%\keywords{Suggested keywords}%Use showkeys class option if keyword
                              %display desired
\maketitle
\textit{Introduction.}---Topological insulators (TIs) 
\cite{kane-mele-z2,kane-mele-qsh,bernevig-prl,bernevig-sci,yang-chang, murakami06,konig,roth,brune,
knez, knez12, fu07, hsieh} 
show anomalous phenomena such as electric conduction on sample surfaces.
Experimentally, the search for candidate materials for TIs is one of the most important problems.
In particular, two-dimensional (2D) TIs are predicted to show unique phenomena, such as the spin Hall effect and 
robust edge states against nonmagnetic impurities, only a few of which have been found \cite{konig,roth,brune,knez,knez12}.
So far, the confirmation of topological materials has been achieved by finding the edge state by 
angle-resolved photoemission spectroscopy or from the transport coefficients.
Since both methods detect anomalous electronic states at the edge,
it is desirable to develop some bulk-sensitive methods that enable us to confirm the topological nature of a material.
In this Letter, we propose that the quantization of the bulk magnetic susceptibility jump
can be used as strong evidence for the topological phase transition in 2D TIs.

%Among bulk physical quantities,
%the magnetic susceptibility can be one that reflects the topological states of a material
%\cite{nakai-nomura, ogata2017}.
Usually, the magnetic susceptibility is discussed in terms of the orbital effect of the magnetic field
\cite{landau,peierls,hebborn1960,blount,hebborn1964, wannier1964,fukuyama,fuku1970,ogata-fukuyama,mcclure,fuku2007,gomez,koshino-ando,ogata3,gao2015,raoux-piechon,
piechon16}
%,xiao2010,thonhauser2011,xiao2010,ceresoli,shi, sundaram, xiao2005, thonhauser2005,ogata2017}
and spin Zeeman effect independently.
In general, however, there can be a cross term between the orbital and Zeeman effects
\cite{yang-chang, murakami06, ito-nomura, loss15,koshino16,nakai-nomura, ogata2017}, 
which we call the orbital--Zeeman (OZ) cross term $\chi_{\rm OZ}$ in the following.
Recently, Nakai and Nomura \cite{nakai-nomura} discussed the jump in $\chi_{\rm OZ}$ at the topological phase transition using the formula 
of the orbital magnetization \cite{xiao2010,thonhauser2011,sundaram,xiao2005,thonhauser2005,ceresoli,shi} 
and the St\v{r}eda formula \cite{streda}.
They calculated the OZ cross term in the Bernevig--Hughes--Zhang model \cite{bernevig-sci} and 
concluded that the width of the jump depends on the $g$-factors of the involved orbitals introduced phenomenologically. 
In general, spin--orbit interaction (SOI) modifies the $g$-factor from its bare value $g_0=2$.
(Here, we neglect the relativistic correction of $g_0$.)
Thus, their conclusion means that the jump in $\chi_{\rm OZ}$ is not quantized in a universal value. 

In the present Letter, we study $\chi_{\rm OZ}$ microscopically based on the Green's function formalism and show that, 
in contrast to the results of Nakai and Nomura, 
the jump in $\chi_{\rm OZ}$ is exactly quantized in units of the universal value $4|e|\mu_{\rm B}/h$ (see Eq.~(\ref{eq:result2}) below)
even in the presence of SOI as long as the $z$-component of the spin is conserved. 
Here, $\mu_{\rm B}=|e|\hbar/2m$ is the Bohr magneton. 
When we study a model microscopically (as in Eq.~(\ref{eq:dirac})), the modification of the $g$-factor does not occur explicitly, 
and instead the effect of SOI appears in the deformation of 
the Bloch wave functions and the energy dispersion, 
which eventually leads to the orbital-dependent $g$-factors.
We show below that the effect of SOI is exactly cancelled out in $\chi_{\rm OZ}$,
which leads to the quantization of jump with a universal value.
We also clarify the physical origin of this result: % of this universal quantization% in $\chi_{\rm OZ}$:
it turns out that the quantization is associated with the chiral edge current, 
which is characteristic of the topological nontrivial state. 
Finally, we apply the obtained formula to Kane--Mele model \cite{kane-mele-qsh, kane-mele-z2} to show the validity of the present proposal. 

\textit{General formalism.}---First, we develop microscopically a general formula for magnetic susceptibility including orbital
magnetism, Pauli paramagnetism, and the OZ cross term in terms of thermal Green's functions in the presence of SOI.
Let $H$ be the general Hamiltonian derived from the 
Dirac equation in the presence of a periodic potential $V(\bm r)$ and a magnetic field, 
which is given by 
\begin{align}
  H=& \frac{1}{2m} ({\bm p} - e{\bm A(\bm r)})^2 - \frac{e\hbar}{2m} \boldsymbol\sigma \cdot {\bm B(\bm r)}
  + V({\bm r}) \nonumber \\
  &+ \frac{\hbar^2}{8m^2c^2} \nabla^2 V 
  + \frac{\hbar}{4m^2c^2} \boldsymbol\sigma \cdot \nabla V \times ({\bm p} -e{\bm A(\bm r)}),\label{eq:dirac}
\end{align}
where ${\bm A(\bm r)}$ is a vector potential, $e<0$ for electrons, 
$\boldsymbol \sigma = (\sigma_x, \sigma_y. \sigma_z)$ are $2\times2$ Pauli matrices,
and ${\bm B(\bm r)}=\nabla \times {\bm A(\bm r)}$ represents a magnetic field.
The last term represents the SOI.
It is to be noted that the second term representing the Zeeman interaction has  
the bare value $g_0=2$.
As performed by Fukuyama \cite{fukuyama}, we implement a perturbative calculation of the free energy in terms of 
the vector potential $\bm A(\bm r)$ %and the magnetic field $\bm B(\bm r)$
via the Luttinger--Kohn representation \cite{lk}.
As a result, we obtain the expression for each contribution as follows 
(The details of the derivation are shown in the Supplemental Material (SM) \cite{sm}.):
\begin{subequations}
	\label{chitotal}
\begin{align}
  &\chi_{\rm orbit}=\frac{e^2}{2\hbar^2} \frac{k_{\rm B} T}{V} \sum_{n\k} 
	{\rm Tr} \,\gamma_x \mathcal{G} \gamma_y \mathcal{G} \gamma_x \mathcal{G} \gamma_y \mathcal{G}, \label{eq:chiorb}\\
  &\chi_{\rm Pauli}=- \frac{k_{\rm B} T}{V} \sum_{n\k} {\rm Tr} \,M_z^s \mathcal{G} M_z^s \mathcal{G}, \\
  &\chi_{\rm OZ}=-\frac{i|e|}{\hbar}  \frac{k_{\rm B} T}{V} \sum_{n\k} 
%	{\rm Tr} \, M_z^s \mathcal{G} (\gamma_x \mathcal{G} \gamma_y \mathcal{G}
%	 -\gamma_y \mathcal{G} \gamma_x \mathcal{G}), \label{chioz}
	{\rm Tr} [M_z^s \mathcal{G} \gamma_x \mathcal{G} \gamma_y \mathcal{G} 
	 -M_z^s \mathcal{G} \gamma_y \mathcal{G} \gamma_x \mathcal{G}],
	\label{chioz}
\end{align}
\end{subequations}
where $\mathcal{G}$ is the thermal Green's function $\mathcal{G}(\k,i\varepsilon_n)$, 
whose $(ll')$ component is the matrix element 
between the $l$th and $l'$th bands.
Each band index includes the pseudo-spin degrees of freedom 
in the case with SOI.
$\varepsilon_n$ is the Matsubara frequency, $\gamma_\mu$ represents
the current operator in the $\mu$-direction divided by $e/\hbar$, 
and $M_z^s$ is the matrix for the operator $-\mu_{\rm B}\sigma_z$.
The effect of SOI is included in $\mathcal{G}$ and $\gamma_\mu$.
Tr is the trace over the band indices and the spin degrees of freedom.
%In Eq. (\ref{eq:chiorb}), $\chi_{\rm orbit}$ is equal to the orbital magnetic susceptibility obtained before \cite{fukuyama} 
%but the effect of SOI is included in $\mathcal{G}$ and $\gamma_\mu$. 
%It is easy to see that $\chi_{\rm Pauli}$ becomes the usual Pauli paramagnetic susceptibility.
%$\chi_{\rm OZ}$ is the OZ cross term.
In Eq.~(\ref{chitotal}), $\chi_{\rm orbit}$ and $\chi_{\rm Pauli}$ respresent the orbital and Pauli magnetic susceptibility, respectively.
These are the same expressions as were obtained before \cite{fukuyama,okuma-ogata15} even in the presence of SOI.
On the other hand, $\chi_{\rm OZ}$ is the OZ cross term, which we focus on in this Letter.

%Now, we rewrite the formula for $\chi_{\rm OZ}$ in Eq.~(\ref{chioz}) in terms of Bloch wave functions
%in a similar way to Ref. \cite{ogata-fukuyama}.
%Note that the total magnetic susceptibility in terms of Bloch wave functions has been obtained before 
%without 
%\cite{hebborn1960,blount,hebborn1964,gao2015, raoux-piechon, piechon16, ogata-fukuyama} and with \cite{ogata2017} the Zeeman interaction. 
%The difference and the relationship with the present formula are discussed later.
To discuss the quantization, we rewrite Eq.~(\ref{chioz}) in terms of the Bloch wave functions 
in a similar way to Ref.~\cite{ogata-fukuyama}.
The periodic part of the Bloch wave function $\hat{u}_{l\k}$ satisfies
\begin{equation}
	H_\k \hat{u}_{l\k}(\bm r) = \varepsilon_l (\bm k) \hat{u}_{l\k}(\bm r),
\end{equation}
where $H_{\k}=e^{-i\k\bm r}He^{i\k\bm r}$ and $\hat{u}_{l\k}(\bm r)$ is a 2-component vector 
$\hat{u}_{l\k}({\bm r})=\,^t(u_{l\k\uparrow}(\bm r), u_{l\k\downarrow}(\bm r))$ in the 
presence of SOI.
%The explicit form of $H_\k$ is shown in the SM \cite{sm}.
In the following, we consider the case where the $z$-component of spin is conserved
even in the presence of SOI.
In this case, up- and down-spin electrons are independent, and 
the energy dispersion is $\varepsilon_{l\sigma}(\k)$ 
(denoted as $\varepsilon_{l\sigma}$ in the following).
The matrices $\mathcal{G}$ and $M_z^s$ are diagonal and given by
  $[\mathcal{G}_\sigma]_{ll'}= 
  \delta_{ll'}(i\varepsilon_n -\varepsilon_{l\sigma} + \mu)^{-1}$ and 
  $[M^s_{z\sigma}]_{ll'} = -\sigma \mu_{\rm B}\delta_{ll'}$, respectively.
On the other hand, the matrix $\gamma_\mu$ has off-diagonal matrix elements between the different bands 
and it becomes,
\begin{align}
  [\gamma_{\mu\sigma}]_{ll'} &= \int u^*_{l\k\sigma} \frac{\partial H_\k}{\partial k_\mu} u_{l'\k\sigma} d{\bm r},
  \nonumber \\
  &=\frac{\partial\varepsilon_{l\sigma}}{\partial k_\mu}\delta_{ll'}
  +(\varepsilon_{l'\sigma}-\varepsilon_{l\sigma})
	\int u_{l\k\sigma}^* \frac{\partial u_{l'\k\sigma}}{\partial k_\mu} d{\bm r}.
\end{align}
where $u_{l\k\sigma}$ is the abbreviation for $u_{l\k\sigma}(\bm r)$.
Substituting these quantities into Eq.~(\ref{chioz}) and carrying out the Matsubara summation, we obtain
\begin{widetext}
\begin{equation}
	\chi_{\rm OZ} = -\frac{2|e|\mu_{\rm B}}{\hbar V} \sum_{l\k\sigma} f(\el)\sigma\Omega_{l\k\sigma}^z 
	+\frac{i|e|\mu_{\rm B}}{\hbar V} \sum_{l\k\sigma} \sigma f'(\varepsilon_l) 
	\left\{ 
		\int  \frac{\partial u^*_{l\k\sigma}}{\partial k_x}(\el-H_\k) \frac{\partial u_{l\k\sigma}}{\partial k_y} d{\bm r}
		-(x \leftrightarrow y) \right\} \; ,
	\label{eq:chiozfull}
\end{equation}
\end{widetext}
where $\Omega_{l\sigma}^z$ is the Berry curvature in the $z$-direction,
\begin{align}
	\Omega_{l\k\sigma}^z=i\int \left( \frac{\partial u^*_{l\k\sigma}}{\partial k_x } \frac{\partial u_{l\k\sigma}}{\partial k_y }
	-\frac{\partial u^*_{l\k\sigma}}{\partial k_y } \frac{\partial u_{l\k\sigma}}{\partial k_x } \right) d{\bm r},
\end{align}
$f(\varepsilon)=(1+e^{(\varepsilon-\mu)/k_{\rm B}T})^{-1}$, 
and the completeness condition $\sum_{l'\sigma}u_{l'\k\sigma}(\bm r) u_{l'\k\sigma}^*(\bm r')=\delta(\bm r-\bm r')$ 
has been used to take the summation over the intermediate state $l'$.

\textit{Universal quantization of} $\chi_{\rm OZ}$.---Let us consider $\chi_{\rm OZ}$ in TIs.
The second term in Eq.~(\ref{eq:chiozfull}) does not contribute in insulators at zero temperature
because the Fermi surface is absent.
Then $\chi_{\rm OZ}$ is written as 
\begin{equation}
	\chi_{\rm OZ} = -\frac{2|e|\mu_{\rm B}}{\hbar V} \sum_{l:{\rm occ}} \sum_{\k\sigma} \sigma \Omega^z_{l\k\sigma},
\end{equation}
where the summation $\sum_{l:{\rm occ}}$ is taken for the occupied bands.
Furthermore, in the case of a 2D insulator, we obtain 
\begin{equation}
	\chi_{\rm OZ}^{\rm 2D}=-\frac{4|e|\mu_{\rm B}}{h}\sum_{l:{\rm occ}}{\rm Ch}_{{\rm s},l}, \label{eq:result2}
\end{equation}
where ${\rm Ch}_{{\rm s},l}$ is the spin Chern number for the $l$th band defined by
\begin{equation}
	{\rm Ch}_{{\rm s},l} = \frac{1}{2}\frac{2\pi}{L^2} \sum_\k (\Omega_{l\k\uparrow} - \Omega_{l\k\downarrow}).
\end{equation}
At a topological phase transition, ${\rm Ch}_{{\rm s},l}$ changes from one integer to another.
Therefore, Eq.~(\ref{eq:result2}) leads to the quantization of the magnetic susceptibility jump at the topological phase transition.
The magnitude of the jump is in units of $\chi_0=4|e|\mu_{\rm B}/h$, which is universal.
Although the effect of SOI is included in $u_{l\k\sigma}$,
it does not affect the coefficient in Eq.~(\ref{eq:result2})
since ${\rm Ch}_{{\rm s},l}$ is a topological number, 
which leads to the universal quantization of jump in $\chi_{\rm OZ}$.

\begin{figure}[h]
	\centering\includegraphics[width=8.6cm]{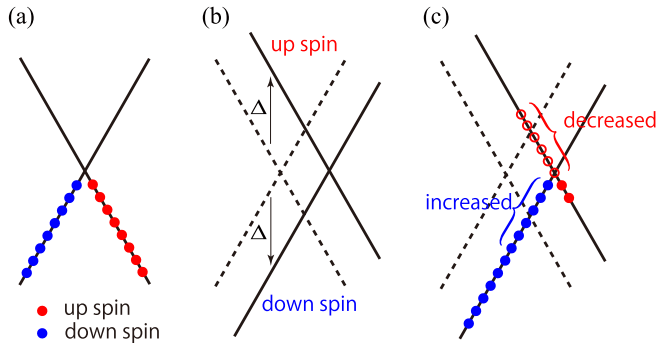}
	\caption{\label{fig:edgedirac} Schematic pictures for the edge state.
	(a) Ground state without an external magnetic field. (b) Change caused by the Zeeman interaction.
	(c) New ground state in the magnetic field.
	}
\end{figure}

\textit{Physical origin of quantization}.---Let us consider the physical origin of this quantization.
$\chi_{\rm OZ}$ is interpreted
as the sum of the correction to the orbital magnetic moment induced by the magnetic field that couples to the spin magnetic moment
and the correction to the spin magnetic moment induced by the magnetic field that couples to the orbital magnetic moment.
Here, we estimate the former correction.
In the edge state, there are spin-polarized linear dispersions and a spin current flows.
Figure \ref{fig:edgedirac}(a) corresponds to the state with ${\rm Ch}_{{\rm s},l}=+1$. 
(Note that the number of pairs of dispersion coincides with $|{\rm Ch}_{{\rm s},l}|$.)
When a magnetic field is applied through the Zeeman interaction $\mu_{\rm B}\sigma_z B$,
the up-spin (down-spin) band moves upward (downward) as shown in Fig. \ref{fig:edgedirac}(b). 
The width of change $\Delta$ is $\mu_{\rm B}B$.
Then, in the lowest energy state [Fig. \ref{fig:edgedirac}(c)], the number of down-spin (up-spin) electrons increases (decreases) 
by $\nu\mu_{\rm B}B$, 
where $\nu$ is the density of states $\nu=L/ch$ and $c$ is the velocity of the edge current.
This change leads to an electric current of $-2|e|\mu_{\rm B}B/h$ in the right direction, 
which causes the orbital magnetic moment of 
$-2|e|\mu_{\rm B}B/h$ per area.
The coefficient of $B$ is half of the quantization of magnetic susceptibility, $-\chi_0/2$.
We can also estimate the contribution from the other correction 
(i.e., the spin magnetic moment induced by an energy shift originating from an orbital magnetic moment 
made by a circular electric current),
which gives the same value.
Combining these two contributions, we obtain the OZ cross term as $-\chi_0$, 
which is consistent with Eq.~(\ref{eq:result2}).

%
%   APPLICATION TO A TOPOLOGICAL INSULATOR
%
\textit{Explicit calculation of} $\chi_{\rm OZ}$ \textit{in Kane--Mele model.}---In the rest of this Letter, 
we calculate the magnetic susceptibility of a model for a 2D TI
to show that $\chi_{\rm OZ}$ actually has a jump at the topological phase transition
and that other contributions do not conceal the quantized jump.
We introduce the Kane--Mele model \cite{kane-mele-z2,kane-mele-qsh,gvv,ljy,ezawanjp,ezawaepj,lfy},
\begin{align}
	H=-&\sum_{\langle i,j \rangle \alpha}t c_{i\alpha}^\dagger c_{j\alpha} 
	+\Delta_0 \left(\sum _{i \in {\rm A}, \alpha} c_{i\alpha}^\dagger c_{i\alpha} 
	- \sum_{i\in {\rm B},\alpha}c_{i\alpha}^\dagger c_{i\alpha} \right ) \nonumber \\
  &+\sum_{\langle\langle i,j \rangle\rangle, \alpha\beta}  i t_2 \nu_{ij}
c_{i\alpha}^\dagger s_{\alpha \beta}^z  c_{j\beta},
\label{tbham}
\end{align}
where $c^\dagger_{i\alpha}$ is the creation operator of an electron 
with spin $\alpha$ at
site $i$, and the summation $\langle i,j \rangle$ ($\langle \langle i,j \rangle \rangle$)
runs over all the nearest- (next-nearest-) neighbor sites of the 2D honeycomb lattice.
The first term represents the usual nearest-neighbor hopping
with transfer integral $t$.
The second term represents a staggered on-site potential,
$+\Delta_0$ for A sublattice and $-\Delta_0$ for B sublattice.
The last term represents the hopping originating from SOI.
We take account of only the $s^z$-component as in Ref.~\cite{kane-mele-qsh} and
set $\nu_{ij}=-\nu_{ji}=+1 \, (-1)$ if the electron makes a left (right) turn to propagate to the 
next-nearest sites.
This model is known as one for silicene 
\cite{gvv,ljy,ezawanjp,ezawaepj,lfy}.
%which is a monolayer of silicon atoms forming a buckled honeycomb lattice.
We can control $\Delta_0$ by changing the electric field 
applied perpendicular to the layer due to its buckled structure.

\begin{figure}[h]
	\centering\includegraphics[width=8.6cm]{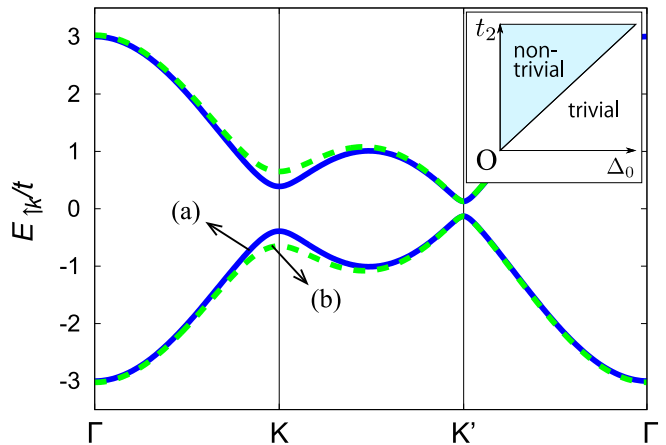}
	\caption{\label{band} Energy dispersion for $\sigma=1$ (up spin) of the model
	in Eq.~(\ref{tbham})
	along the path $\Gamma \rightarrow  K \rightarrow  K' \rightarrow \Gamma$
	for two typical choices of parameters:
	(a) solid line, $\Delta_0/t=1/4,t_2/t=\sqrt{3}/36$ (topologically trivial)
	and (b) dashed line, $\Delta_0/t=1/4,t_2/t=\sqrt{3}/12$ (topologically nontrivial).
	The energy dispersion for $\sigma=-1$ (down spin) is obtained by exchanging $K$ for $K'$ points.
	Inset:
	Phase diagram of this model. 
	}
\end{figure}

The energy dispersion of this model
is shown in Fig.~\ref{band} for case (a) with $\Delta_0/t=1/4,t_2/t=\sqrt{3}/36$
(topologically trivial; solid line) and for case (b) with $\Delta_0/t=1/4,t_2/t=\sqrt{3}/12$ 
(topologically nontrivial; dashed line).
In the following, we use (a) and (b) as typical cases.
The explicit expression of the 
energy dispersion is shown in SM \cite{sm}.
Since the space inversion symmetry is broken,
the energy dispersions for up and down spins can be different.
In the momentum space, gaps open at $K=(4\pi/3\sqrt{3}a,0)$ and $K'=(-4\pi/3\sqrt{3}a,0)$
with $a$ being the distance between the nearest-neighbor sites.
Their magnitudes are $2\left| \Delta_0 + (3\sqrt{3}/2) \sigma t_2\right|$ at $K$ and
$2\left| \Delta_0 - (3\sqrt{3}/2) \sigma t_2\right|$ at $K'$, respectively,
where $\sigma=1$ is for up spin and $\sigma=-1$ is for down spin.

In this model, the ratio of $\Delta_0$ to $t_2$ 
determines the topological order \cite{lfy,ezawanjp,ezawaepj}:
topologically trivial for $|t_2/\Delta_0|<2/3\sqrt{3}$ and 
topologically nontrivial for $|t_2/\Delta_0|> 2/3\sqrt{3}$.
The phase diagram is shown in the inset of Fig. \ref{band}.

\begin{figure}
%\centering\includegraphics[width=8cm]{berry4.eps}
\centering\includegraphics[width=8.6cm]{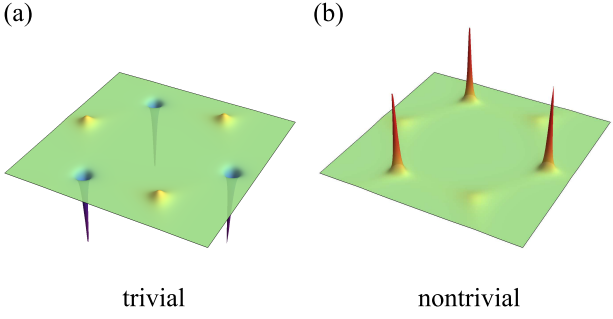}
\caption{\label{berry} 
Distribution of Berry curvature in the momentum space 
for the valence band electrons with up spin
for case of (a) (topologically trivial) 
and (b) (topologically nontrivial).
The Chern numbers are $0$ for (a) and $1$ for (b).
}
\end{figure}

Before calculating magnetic susceptibility, let us examine the Berry curvature.
Figure \ref{berry} shows the distribution of the Berry curvature in the momentum space for the valence
band electrons with up spin
for the two choices of the parameters in Fig.~\ref{band}. 
It is seen that the Berry curvature is localized near $K$ and $K'$ points.
After numerical integration, we find that the Chern numbers are $0$ for (a)
and $1$ for (b),
which is consistent with the fact that cases (a) and (b)
belong to the topologically trivial and nontrivial phases, respectively.
According to Figs.~\ref{band} and \ref{berry}, 
the low-energy excitations in the vicinity of $K$ and $K'$ points are important when $\mu\simeq 0$.
Therefore, we approximate the Hamiltonian
by the expansion around $K$ and $K'$ points, i.e., $\bm k \cdot \bm p$ perturbation.
In this way, we obtain a low-energy effective model,
\begin{align}
h^{K/K'}_{\sigma} = (\Delta^{K/K'}_\sigma -\sigma \alpha k^2 )\tau_z 
+ \hbar v_{\rm F} k_x \tau_x + \hbar v_{\rm F} k_y \tau_y, \label{effham}
\end{align}
where 
$\tau_x$, $\tau_y$, and $\tau_z$ are the Pauli matrices representing the degrees of 
freedom of sublattices A and B,
$\Delta^{K}_\sigma=\Delta_0 + \sigma(3\sqrt{3}/2) t_2$,
$\Delta^{K'}_\sigma=-\Delta_0 + \sigma(3\sqrt{3}/2) t_2$, 
$\alpha=(9\sqrt{3}/8)a^2 t_2$, 
and $\hbar v_{\rm F}=(3/2)at$.
Note that the signs of the mass term $\Delta_\sigma^{K/K'}$ at $K$ and $K'$ points 
with different spins are opposite, i.e.,
$\Delta^K_\uparrow=-\Delta^{K'}_\downarrow$ and $\Delta^K_\downarrow=-\Delta^{K'}_\uparrow$.
This effective Hamiltonian is justified in the limit of 
$t_2,\Delta_0 \rightarrow 0$ with $t_2/\Delta_0$ fixed.
%\begin{figure}[th]
%	\centering\includegraphics[width=8.0cm]{chi-full3.eps}
%	\caption{\label{chimu1} 
%		Chemical potential $\mu$ dependences of each contribution to magnetic susceptibility 
%		for the case (a) (topologically trivial) and (b) (topologically nontrivial).
%		Each contribution is normalized as 
%		$\tilde{\chi}_{\rm OZ}=\chi_{\rm OZ}/\chi_0$, $\tilde{\chi}_{\rm orbit}=\chi_{\rm orbit}/C_1\chi_0$, 
%		and $\tilde{\chi}_{\rm Pauli}=\chi_{\rm Pauli}C_1/\chi_0$ with 
%		$C_1=(\frac{1}{2}mv_{\rm F}^2)/t$.
%		}
%\end{figure}

Let us discuss the magnetic susceptibility.
Note that Ezawa \cite{ezawaepj} calculated the orbital magnetism but the OZ cross term was not taken into account.
In the model Eq.~(\ref{effham}), the thermal Green's function is defined as 
$\mathcal{G}_\sigma^{K/K'}=(i\varepsilon_n - h^{K/K'}_\sigma)^{-1}$.
The current operator in the $\mu$-direction is given by 
$\gamma_\mu= 	\hbar v_{\rm F} \tau_\mu - 2 \sigma \alpha k_\mu \tau_z$.
Substituting these quantities into Eq.~(\ref{chitotal}),
carrying out the Matsubara summation, and performing the 2D momentum integration
at $T=0$, we obtain
\begin{subequations}
	\label{eq:chigrares}
\begin{align}
	&\chi_{\rm orbit}=-\frac{e^2v_{\rm F}^2}{6\pi}
	\sum_{\eta=K,K'} \frac{1}{|\Delta_\uparrow^\eta|} 
	\theta (|\Delta_\uparrow^\eta| - |\mu|), 
	\label{chiorbit2}\\
	&\chi_{\rm Pauli}=\frac{\mu_{\rm B}^2}{\pi \hbar^2v_{\rm F}^2} 
	\sum_{\eta=K,K'} |\mu|\theta ( |\mu|- |\Delta_\uparrow^\eta| ),
	\label{chipauli2}\\
	&\chi_{\rm OZ} =- \frac{2\mu_{\rm B}|e|}{h} 
    \sum_{\eta=K,K'}{\rm sgn} (\Delta_\uparrow^\eta )\theta ( |\Delta_\uparrow^\eta |-|\mu|),
%	&\qquad\qquad \qquad\qquad \left. -{\rm sgn}\left(\Delta_\downarrow^K\right)
%	\theta\left(\left|\Delta_\downarrow^K\right|-|\mu|\right)\right],
	\label{chicross2}
\end{align}
\end{subequations}
in the limit of $t_2, \Delta_0 \rightarrow 0$ with $t_2/\Delta_0$ fixed.
Here, $\chi_{\rm orbit}$ is the orbital diamagnetic susceptibility of the
2D Dirac electrons discussed in the preceding studies 
\cite{mcclure,fuku2007,gomez,ogata3,koshino-ando,gao2015,raoux-piechon}.
$\chi_{\rm Pauli}$ is the Pauli paramagnetism proportional to the density of states ($\propto |\mu|$).

\begin{figure}[th]
	\centering\includegraphics[width=8.6cm]{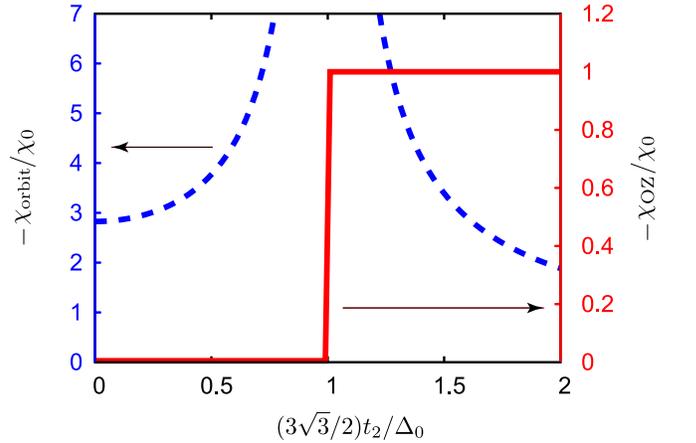}
	\caption{\label{figchioz}
	Contributions to magnetic susceptibility as a function of 
	$(3\sqrt{3}/2)t_2/\Delta_0$ for $\mu=0$ with $\Delta_0=t/4$.
	%They are normalized as 
	%$\tilde{\chi}_{\rm OZ}=\chi_{\rm OZ}/\chi_0$ and 
	%$\tilde{\tilde{\chi}}_{\rm orbit}=\chi_{\rm orbit}/C_2\chi_0$
	%and $\tilde{\tilde{\chi}}_{\rm Pauli}=\chi_{\rm Pauli}\chi_0$
	%with $C_2=(\frac{1}{2}mv_{\rm F}^2)/\Delta_0$.
	%Each contributions are in different units from Fig.~\ref{chimu1}.
	%Normalization for $\chi_{\rm orbit}$ is different from that in Fig.~\ref{chimu1}.
	They are normalized by $\chi_0$.
	The values of $v_{\rm F}$ and $a$ are chosen to be the same as those of graphene.
	The system is topologically trivial (nontrivial) in the region $(3\sqrt{3}/2)t_2/\Delta_0 <1 (>1)$.
	}
\end{figure}

To observe the quantization of the jump, we focus on the case of $\mu=0$, an insulating case, 
where $\chi_{\rm Pauli}$ vanishes.
Figure~\ref{figchioz} shows $\chi_{\rm orbit}$ and $\chi_{\rm OZ}$ as a function of 
$t_2 /\Delta_0$.
When $(3\sqrt{3}/2)t_2/\Delta_0 <1$,
the system is topologically trivial and 
the signs of $\Delta_\uparrow^K$ and $\Delta_\uparrow^{K'}$ are opposite,
which leads to $\chi_{\rm OZ}=0$ from Eq.~(\ref{chicross2}).
When $(3\sqrt{3}/2)t_2/\Delta_0 >1$, on the other hand, the system is topologically nontrivial and
the signs of $\Delta_\uparrow^K$ and $\Delta_\uparrow^{K'}$ are the same, which leads to $\chi_{\rm OZ}=-\chi_0$.
As a result, $\chi_{\rm OZ}$ has a universal jump at the topological phase transition 
at $t_2/\Delta_0=2/3\sqrt{3}$.
On the other hand, $\chi_{\rm orbit}$ diverges at the phase transition due to the gap closing.
However, we can see that 
$\chi_{\rm orbit}=-\frac{e^2 v_{\rm F}^2}{6\pi}(|\Delta_0-3\sqrt{3}t_2/2|^{-1} + |\Delta_0+3\sqrt{3}t_2/2|^{-1})$
and the magnitude of divergence is the same on both sides of the phase transition.
Therefore, when we subtract the divergence of $\chi_{\rm orbit}$, we will be able to detect the jump in $\chi_{\rm OZ}$.
%As shown in Fig.~\ref{figchioz}, the magnitude of $\chi_{\rm OZ}$ is comparable to
%that of $\chi_{\rm orbit}$ when $\Delta_0=t/4$.
Note that the effect of SOI represented by $t_2$ appears only in the magnitude of $\chi_{\rm orbit}$ 
and does not affect the magnitude of $\chi_{\rm OZ}$.

\textit{Discussion and conclusion.}---The expression for the magnetic susceptibility including the spin Zeeman effect
was obtained in terms of Bloch wave functions 
\cite{ogata2017},
which contains a term
$\chi_{\rm occ:2} = -\frac{e}{\hbar} {\rm Re} \sum_{l\k} f(\varepsilon_\k) M_{ll}^z \Omega_l^z.$
Actually, $\chi_{\rm occ:2}$ gives half of $\chi_{\rm OZ}$ in Eq.~(\ref{eq:result2}).
The origin of this difference is as follows.
In Ref.~\cite{ogata2017}, the effect of Zeeman interaction is distributed among several terms 
for total magnetic susceptibility including $\chi_{\rm occ:2}$.
Therefore, if we collect all the effects of Zeeman interaction in the formalism of Ref.~\cite{ogata2017},
we can recover $\chi_{\rm OZ}$.

Based on the microscopic theory, we have derived a new simple formula for magnetic susceptibility 
in a Bloch system with SOI and Zeeman interaction
to show that the OZ cross term, one of the three contributions to magnetic susceptibility, 
is always quantized in units of the universal value $4|e|\mu_{\rm B}/h$ for 2D spin-conserving insulators at zero temperature.
%This quantization originates from the magnetic response of the chiral edge state.
We have clarified that this quantization originates from the redistribution
of the chiral edge state due to the magnetic field.
We have also applied the formula to a model for a 2D TI and 
demonstrated the quantization.
Our results clearly show that the magnetic response reflects the topological nature of a material.
It should be possible, therefore, to make a bulk-sensitive confirmation of the topological phase transition.

\begin{acknowledgments}
We thank very fruitful discussions with H.\ Matusura, H.\ Maebashi, I.\ Tateishi, T.\ Hirosawa, N.\ Okuma, and V. K\"onye.
This work was supported by Grants-in-Aid for Scientific Research from the Japan Society for the Promotion of Science
(Grants No.~JP18H01162).
S.O. was supported by the Japan Society for the Promotion of Science through the Program for Leading Graduate Schools
(MERIT).
\end{acknowledgments}
%\nocite{*}
\bibliography{crossresp_v10}% Produces the bibliography via BibTeX.

\end{document}

\preprint{APS/123-QED}

\title{Supplemental Material}% Force line breaks with \\\thanks{A footnote to the article title}%

\author{Soshun Ozaki}
% \altaffiliation[Also at ]{Physics Department, XYZ University.}%Lines break automatically or can be forced with \\
\author{Masao Ogata}%
% \email{Second.Author@institution.edu}
\affiliation{%
 Department of Physics, University of Tokyo, Bunkyo, Tokyo 113-0033, Japan
% Authors' institution and/or address\\
% This line break forced with \textbackslash\textbackslash
}

\date{\today}
\maketitle

%\tableofcontents
%The system with a periodic potential $V({\bm r})$ in presence of magnetic field 
%is goverened by the Hamiltonian
\onecolumngrid
\section{derivation of the formula for magnetic susceptibility}
We consider Bloch electrons in a periodic potential $V({\bm r})$,
Zeeman interaction, and SOI.
For this purpose, we use the following Hamiltonian derived from the Dirac equation
in the presence of a magnetic field:
\begin{align}
  H= \frac{1}{2m} ({\bm p} - e{\bm A(\bm r)})^2 - \frac{e\hbar}{2m} \sigma \cdot {\bm B(\bm r)}
  + V({\bm r})
  + \frac{\hbar^2}{8m^2c^2} \nabla^2 V 
  + \frac{\hbar}{4m^2c^2} \boldsymbol\sigma \cdot \nabla V \times ({\bm p} -e{\bm A(\bm r)}),
\end{align}
where ${\bm A(\bm r)}$ is a vector potential, $e<0$ for electrons, 
and ${\bm B(\bm r)}=\nabla \times {\bm A(\bm r)}$ represents a magnetic field.
The $g$-factor for electrons is assumed to be $g=2$.
%We estimate the thermodynamic potential by a second-order perturbation 
%with respect to ${\bm A}$ and ${\bm B}$.

The magnetic susceptibility per volume is defined as 
\begin{align}
  \chi=- \frac{1}{V} \lim_{B \rightarrow 0} \frac{\partial^2 \Omega}{\partial B^2},
\end{align}
where $\Omega$ is the thermodynamic potential of the system with chemical potential $\mu$, i.e.,
\begin{align}
  \Omega = -k_{\rm B} T \ln {\rm Tr} \exp [-\beta(H-\mu N)].
\end{align}
Therefore, it is sufficient to evaluate the second-order deviation of $\Omega$ due to 
a magnetic field, defined as $\Omega^{(2)}$.

To perform the pertubative calculations with respect to ${\bm A}$ and ${\bm B}$
as was done by Fukuyama \cite{fukuyama}, we rewrite
the Hamiltonian in a second quantized form as 
\begin{align}
  H=H_0 + H_1 + H_2
\end{align}
with
\begin{align}
  &H_0= \sum_{\alpha\beta} \int \psi^\dagger_\alpha({\bm r}) 
	\left\{ -\frac{\hbar^2 }{2m}\nabla^2 +V({\bm r}) + \frac{\hbar^2}{8m^2c^2}\nabla^2 V({\bm r}) 
	+ \frac{\hbar}{4m^2c^2} \boldsymbol \sigma_{\alpha\beta} \cdot \nabla V \times {\bm p} \right\} \psi_\beta({\bm r}) d{\bm r} \nonumber \\
  &H_1 = -\int {\bm j}({\bm r}) \cdot {\bm A}({\bm r}) d{\bm r} 
  - \sum_{\alpha\beta} \int \psi^\dagger_\alpha ({\bm r}) \frac{e\hbar}{2m} \boldsymbol \sigma_{\alpha\beta} \cdot {\bm B(\bm r)} 
  \psi_\beta({\bm r}) d{\bm r} \nonumber \\
  &H_2 = \sum_\alpha \frac{e^2}{2m} \int {\bm A}^2({\bm r}) \psi^\dagger_\alpha({\bm r}) \psi_{\beta}({\bm r}) d{\bm r},
\end{align}
where ${\bm j}({\bm r})$ is the current operator without a vector potential,
\begin{align}
  {\bm j} = \frac{e}{2m} \sum_\alpha \{ \psi^\dagger_\alpha( {\bm p} \psi_\alpha) -({\bm p}\psi^\dagger_\alpha)\psi_\alpha \}
  + \frac{e\hbar}{4m^2c^2}\sum_{\alpha\beta}\psi^\dagger_\alpha \boldsymbol \sigma_{\alpha\beta} \times \nabla V  \psi_\beta.
\end{align}
The subscript of each Hamiltonian represents the order of $\bm A$.
To avoid unphysical contributions due to the unbounded character of the 
coordinate operators, we introduce the vector potential $\bm A$ with 
Fourier component ${\bm q}$, as was done by Hebborn and Sondheimer \cite{hebborn1960},
\begin{align}
  {\bm A}({\bm r}) = -i {\bm A}_{\bm q} 
  (e^{i {\bm q \bm r}} - e^{-i{\bm q \bm r}}),
\end{align}
and let ${\bm q}=0$ in the final expressions.

To make the pertubative calculation easier, we employ the Luttinger--Kohn representation \cite{lk},
\begin{align}
  \psi_\alpha({\bm r})= \sum_{l \k \alpha} \chi_{l \k\alpha}(\bm r) a_{l\k\alpha}, \quad 
  \chi_{l\k\alpha}({\bm r})= e^{i{\bm k \bm r}}u_{l \k_0 \alpha} ({\bm r}),
\end{align}
where $\k_0$ is a fixed wave vector and $\hat{u}_{l\k}(\bm r)=\,^t\!\left(u_{l\k\uparrow}(\bm r), u_{l\k\downarrow}(\bm r)\right)$ 
is the periodic part of the Bloch wave function with the wave vector $\k$ of the $l$th band.
$\hat{u}_{l\k}$ satisfies 
\begin{equation}
  H_\k \hat{u}_{l\k} (\bm r) =\varepsilon_{l\k} \hat{u}_{l\k}(\bm r)
\end{equation}
with
\begin{align}
  H_\k = \frac{\hbar^2}{2m} (\k- i\nabla)^2 + V(\bm r) + \frac{\hbar^2}{8m^2c^2}\nabla^2 V
  + \frac{\hbar^2}{4m^2c^2} \boldsymbol \sigma \cdot \nabla V \times (\k -i\nabla).
\end{align}
In this representation, $H_0$, $H_1$, and $H_2$ are rewritten as 
%\onecolumngrid
\begin{align}
  H_0=&\sum_{ll'\alpha\beta \k} E_{ll'\alpha\beta} a_{l\k\alpha}^\dagger a_{l'\k\beta}, 
  \nonumber \\
  H_1=&\frac{ie}{\hbar} A_{\q \mu}\sum_{ll'\alpha\beta \k} \gamma_{\mu}(\k)
  (a_{l\k+\frac{\q}{2}\alpha}^\dagger a_{l'\k-\frac{\q}{2}\beta}
  -a_{l\k-\frac{\q}{2}\alpha}^\dagger a_{l'\k+\frac{\q}{2}\beta}) \nonumber \\
  &-\q \times {\bm A}_\q \cdot 
  \sum_{ll'\alpha\beta \k} {\bm M}^s 
  (a_{l\k+\frac{\q}{2}\alpha}^\dagger a_{l'\k-\frac{\q}{2}\beta}
  +a_{l\k-\frac{\q}{2}\alpha}^\dagger a_{l'\k+\frac{\q}{2}\beta}), \nonumber \\
  H_2=&-\frac{e^2}{2m}\sum_\alpha {\bm A}_\q^2 (\rho_{-2\q\alpha}-2\rho_{0\alpha}+\rho_{2\q\alpha}),
\end{align}
where
\begin{align}
	&E_{ll'\alpha\beta}=\int u_{l\k_0 \alpha}^* H_{\k} u_{l'\k_0 \beta} d{\bm r}, \\
	&[\gamma(\k)]_{l\alpha,l'\beta}=\int u_{l\k_0 \alpha}^* \frac{\partial H_{\k}}{\partial \k} u_{l'\k_0\beta}d{\bm r}, \\
	&[{\bm M^s}]_{l\alpha,l'\beta}= \int u_{l\k_0\alpha}^* \frac{e\hbar}{2m} \boldsymbol \sigma u_{l'\k_0\beta} d{\bm r}, \\
	&\rho_\q = \sum_{l\k\alpha} a_{l\k-\q\alpha}^\dagger a_{l\k\alpha}.
\end{align}
%\twocolumngrid

\begin{figure*}[tb]
\includegraphics[width=0.8\linewidth]{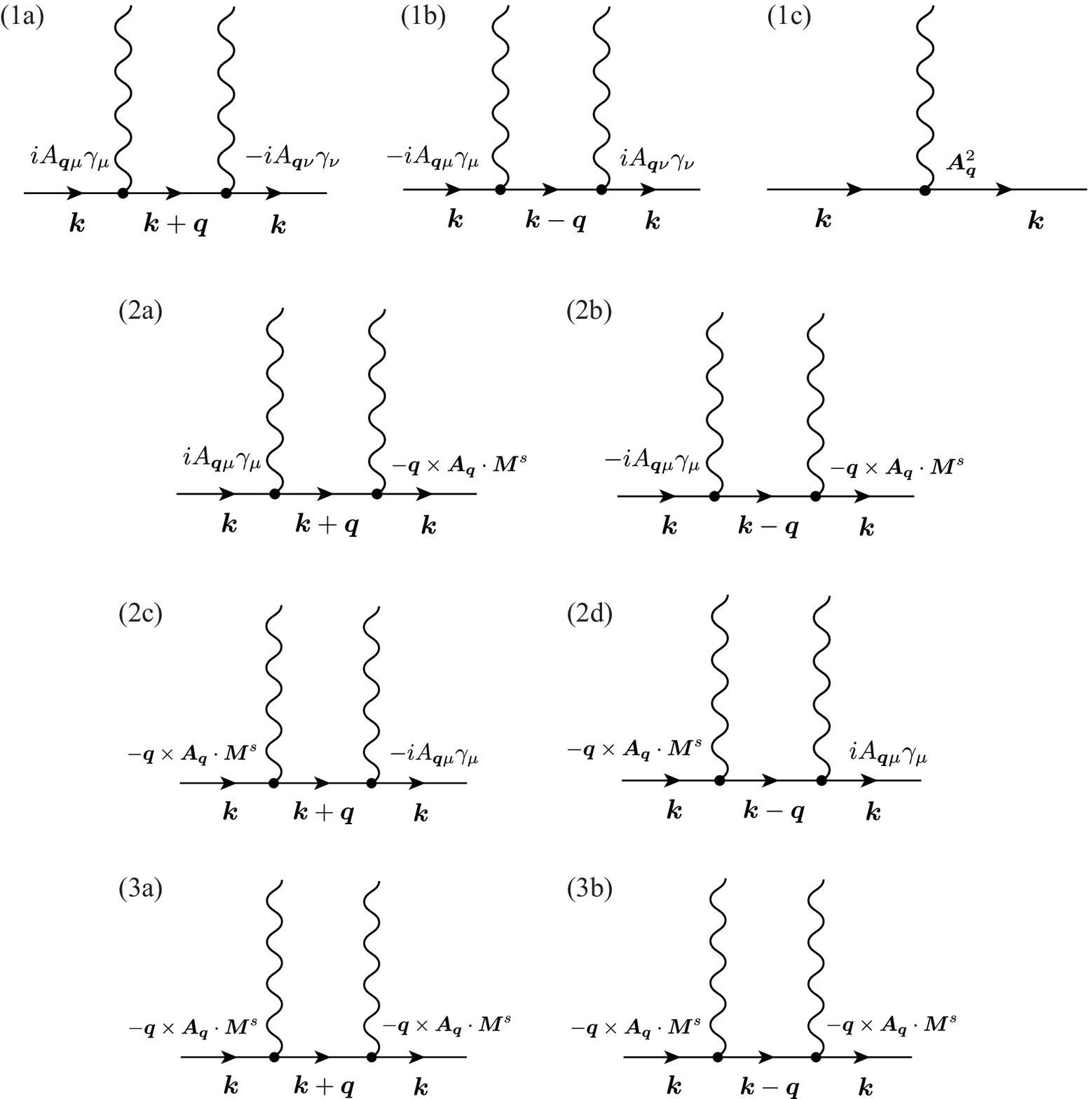}
\caption{\label{diagrams} Feynman diagrams for the second-order deviation of the free energy $\Omega^{(2)}$.}
\end{figure*}

By using the Luttinger--Kohn representation, the thermodynamic potential becomes
\begin{align}
  \Omega=\Omega_0 + k_B T \int_0^1 \frac{d\lambda}{\lambda} 
  \sum_{{\bm k}n} {\rm Tr} 
  \left[ \Sigma^{\star \lambda}(\k,\varepsilon_n) \mathcal{G}^\lambda (\k,\varepsilon_n) \right],
\end{align}
where $\Omega_0$ is the thermodynamic potential without a vector potential.
$\Sigma^{\star\lambda}$ and $\mathcal{G}^\lambda$ are the self-energy matrix and thermal Green's function 
corresponding to the Hamiltonian $H_0 + \lambda(H_1+H_2)$, respectively.
We expand the self-energy as 
\begin{align}
	\Sigma^{\star \lambda}=\lambda \Sigma^{\star (1)} + \lambda^2 \Sigma^{\star (2)} + O(A^3).
\end{align}
Making use of Dyson's equation,
\begin{align}
	\mathcal{G}^\lambda(\k,i\varepsilon_n)
	=\mathcal{G}^{(0)}(\k,i\varepsilon_n) + \mathcal{G}^{(0)}(\k,i\varepsilon_n)\Sigma^{\star\lambda}\mathcal{G}^\lambda(\k,i\varepsilon_n),
\end{align}
with $\mathcal{G}^{(0)}$ being Green's function for the nonperturbative Hamiltonian $H_0$,
we can carry out the integral as 
\begin{align}
	\Omega-\Omega_0=k_{\rm B} T \sum_{\k n} {\rm Tr} \left[\Sigma^{\star(1)}\mathcal{G}^{(0)}+ \frac{1}{2} \Sigma^{\star(2)}\mathcal{G}^{(0)} \right]
	+O(A^3). \label{deviation}
\end{align}
To estimate Eq. (\ref{deviation}), we find nine types of contributions of the order of ${\bm A^2}$ 
, whose diagrams are shown in Fig. 1.
Figures \ref{diagrams}(1a)--(1c), \ref{diagrams}(2a)--(2d), and \ref{diagrams}(3a) and \ref{diagrams}(3b)
give the orbital contribution, the orbital-Zeeman cross term contribution, 
and the Zeeman contribution to the free energy, respectively.
Each contribution is calculated as
%\onecolumngrid
\begin{align}
	\Omega^{(2)}_{\rm orbit}=&\frac{1}{2} k_{\rm B} T \sum_{n\k} \frac{e^2}{\hbar^2} A_{\q\mu}A_{\q\nu} 
	\tr \left[ \gamma_\mu \mathcal{G}(\k+\frac{\q}{2},i\varepsilon_n)
	\gamma_\nu \mathcal{G}(\k -\frac{\q}{2},i\varepsilon_n) + (\q \leftrightarrow -\q) \right] \nonumber \\
	&+ k_{\rm B}T \sum_{n\k}\frac{e^2}{m} {\bm A}_{\bm q}^2 {\rm Tr} \mathcal{G}({\bm k},i\varepsilon_n), \\
\nonumber \\
	\Omega^{(2)}_{\rm OZ}=
	&-\frac{1}{2} k_{\rm B} T \sum_{n\k} \tr[ \frac{ie}{\hbar}A_{\q\mu} \gamma_\mu \mathcal{G}(\k+\frac{\q}{2}, i\varepsilon_n) 
	\q \times {\bm A_q} \cdot {\bm M}^s \mathcal{G}(\k-\frac{\q}{2},i\varepsilon) +(\q \rightarrow -\q)] \\
	&+\frac{1}{2} k_{\rm B} T \sum_{n\k} \tr [\q \times {\bm A_q} \cdot {\bm M}^s
  \mathcal{G}(\k+\frac{\q}{2},i\varepsilon_n) \frac{ie}{\hbar} A_{\q \mu}\gamma_\mu \mathcal{G}(\k-\frac{\q}{2},i\varepsilon_n) 
  +(\q \leftrightarrow -\q)], \nonumber \\
\nonumber \\
	\Omega^{(2)}_{\rm Zeeman}=&\frac{1}{2} k_{\rm B} T \sum_{n\k} 
	\tr [\q \times {\bm A_q} \cdot {\bm M}^s \mathcal{G}(\k+\frac{\q}{2},i\varepsilon_n)
	\q \times {\bm A_q} \cdot {\bm M}^s \mathcal{G}(\k-\frac{\q}{2},i\varepsilon_n) + (\q \leftrightarrow -\q)],
\end{align}
respectively.
%\twocolumngrid
By expanding the expression for small $\q$
with the help of the Ward identity,
\begin{align}
\mathcal{G}(\k,i\varepsilon_n)\gamma_\mu \mathcal{G}(\k,i\varepsilon_n)
= \frac{\partial \mathcal{G}(\k, i\varepsilon_n)}{\partial k_\mu},
\end{align}
we obtain
%Eq. (11) can be written in the limit of small $\bm q$ as
\begin{align}
  \Omega^{(2)}_{\rm orbit}=&-\frac{e^2}{2\hbar^2}({\bm q} \times A_{\bm q})_z^2 k_B T 
  \sum_{n\k} {\rm Tr} \gamma_x \mathcal{G} \gamma_y \mathcal{G} \gamma_x \mathcal{G} \gamma_y \mathcal{G} \nonumber \\
  \Omega^{(2)}_{\rm OZ}=&-\frac{ie}{\hbar} ({\bm q} \times A_{\bm q})_z^2 k_B T
  \sum_{n\k} {\rm Tr} [M_z^s \mathcal{G} \gamma_x \mathcal{G} \gamma_y \mathcal{G}
	-M_z^s \mathcal{G} \gamma_y \mathcal{G} \gamma_x \mathcal{G}] \nonumber \\
  \Omega^{(2)}_{\rm Zeeman}=&({\bm q} \times A_{\bm q})_z^2 k_B T 
  \sum_{n\k} {\rm Tr} M_z^s \mathcal{G} M_z^s \mathcal{G}.
\end{align}
Since $2({\bm q} \times A_{\bm q})_z^2$ corresponds to $B_z^2$,
the magnetic susceptibility becomes
\begin{align}
  \chi=\chi_{\rm orbit} + \chi_{\rm OZ} + \chi_{\rm Pauli},
\end{align}
where
\begin{align}
  &\chi_{\rm orbit}=\frac{e^2}{2\hbar^2} \frac{k_{\rm B} T}{V} \sum_{n\k} 
	{\rm Tr} \gamma_x \mathcal{G} \gamma_y \mathcal{G} \gamma_x \mathcal{G} \gamma_y \mathcal{G} \nonumber \\
  &\chi_{\rm OZ}=\frac{ie}{\hbar} \frac{k_{\rm B} T}{V} \sum_{n\k} {\rm Tr} 
	[M_z^s \mathcal{G} \gamma_x \mathcal{G} \gamma_y \mathcal{G}
	-M_z^s \mathcal{G} \gamma_y \mathcal{G} \gamma_x \mathcal{G}]
	\nonumber \\
  &\chi_{\rm Pauli}=- \frac{k_{\rm B} T}{V} \sum_{n\k} {\rm Tr} M_z^s \mathcal{G} M_z^s \mathcal{G}. \label{result}
\end{align}
%Here, $\chi_{\rm Pauli}$ is derived from $\Omega^{(2)}_{\rm Zeeman}$
%and is an extention of Pauli paramgnetism.
Since the Luttinger--Kohn representation is associated with the Bloch representation
by a unitary transformation \cite{lk}, the expressions in Eq. (\ref{result}) can be regarded as written in terms of 
Green's functions and current operators for the Bloch representation.

\section{energy dispersion of the model}
Performing the Fourier transform, we obtain the Hamiltonian in a matrix form,
\begin{eqnarray}
	H_{\k\sigma} = \left(
		\begin{array}{cc}
			\Delta_0 + \sigma \lambda_\k & -t \gamma_\k^* \\
			-t \gamma_\k & -\Delta_0 - \sigma \lambda_\k
		\end{array}
		\right), \label{ham2}
\end{eqnarray}
where $\sigma=1$ is for up spin and $\sigma=-1$ is for down spin.
The complex factor $\gamma_\k$,
\begin{align}
	\gamma_\k= e^{-ik_y a} + e^{i (\frac{\sqrt{3}}{2}k_x + \frac{1}{2}k_y )a}
		+e^{i (-\frac{\sqrt{3}}{2}k_x + \frac{1}{2}k_y )a},
\end{align}
comes from the Fourier transform of hopping
with $a$ being the distance between the nearest-neighbor sites, and
\begin{align}
	\lambda_\k = 2 t_2 \sin\frac{\sqrt{3}}{2} k_x a
		\left(\cos\frac{3}{2}k_y a - \cos\frac{\sqrt{3}}{2} k_x a \right)
\end{align}
comes from the spin-dependent hopping, $t_2$.
The energy dispersion of this model is given by
\begin{equation}
  E^\pm_{\sigma\bm k}=\pm \sqrt{(\Delta_0+\sigma\lambda_{\bm k})^2 + t^2 |\gamma_\k|^2}.
\end{equation}

\bibliography{crossresp_v10}% Produces the bibliography via BibTeX.